\documentstyle[prc,aps,epsfig]{revtex}
\newcommand{\gsim}{\mathrel{\hbox{\rlap{\lower.55ex \hbox {$\sim$}}
                   \kern-.3em \raise.4ex \hbox{$>$}}}}
\newcommand{\lsim}{\mathrel{\hbox{\rlap{\lower.55ex \hbox {$\sim$}}
                   \kern-.3em \raise.4ex \hbox{$<$}}}}
\newcommand{\beq}{\begin{equation}}
\newcommand{\eeq}{\end{equation}}
\newcommand{\bea}{\begin{eqnarray}}
\newcommand{\eea}{\end{eqnarray}}
\newcommand{\ba}{\begin{array}}
\newcommand{\ea}{\end{array}}

\newcommand{\etal}{{\it et al.}}
\newcommand{\eg}{{\it e.g.}}
\newcommand{\ie}{{\it i.e.}}
\newcommand{\rhs}{{\it rhs}}

\newcommand{\wrt}{{\it wrt}}
\def\la{\langle}
\def\ra{\rangle}

\newcommand{\bef}{\begin{figure}}
\newcommand{\eef}{\end{figure}}
\newcommand{\bce}{\begin{center}}
\newcommand{\ece}{\end{center}}

\begin{document}

\tightenlines

\title{A chiral crystal in cold QCD matter at intermediate densities?}

\author{Ralf Rapp, Edward Shuryak and Ismail Zahed}

\address{Department of Physics and Astronomy, State
University of New York, Stony Brook, NY 11794-3800, USA}

\maketitle

\begin{abstract}
The analogue of Overhauser (particle-hole) pairing in electronic systems
(spin-density waves with non-zero total momentum $Q$) 
is analyzed in finite-density QCD for 3 colors and 
2 flavors, and compared to the color-superconducting BCS ground state 
(particle-particle pairing, $Q$=0).
The calculations are based on effective nonperturbative 
four-fermion interactions acting in both the scalar diquark as well
as the scalar-isoscalar quark-hole ('$\sigma$') channel. 
Within the Nambu-Gorkov formalism we set up the  
coupled channel problem including multiple
chiral density wave formation, and evaluate 
the resulting gaps and free energies. 
Employing medium-modified instanton-induced
't Hooft interactions, as applicable around $\mu_q\simeq 0.4$~GeV (or 
4 times nuclear saturation density),  we find the 'chiral crystal phase'  
to be competitive with the color superconductor.

\end{abstract}


\section{Introduction}
\label{sec_intro}
The understanding of QCD under extreme conditions is among  the
main frontiers in strong interaction physics. In particular, the
finite-density and zero-temperature regime has reattracted considerable
attention lately,
after it has been realized that early perturbative estimates for 
color-superconducting gaps at large chemical
potential are exceeded  by up to two orders of magnitude towards smaller 
densities~\cite{ARW98,RSSV98,ARW99,RSSV00}. Such BCS-type pairing energies  
are in fact comparable to the ('constituent') quark mass gap in 
the QCD vacuum,  $M_q\simeq$~0.35-0.4~GeV, and have triggered new 
interest in observable consequences of    
quark matter formation within the core of neutron stars 
(unfortunately, in high energy heavy-ion collision large entropy 
production renders the access to this regime unlikely). 

The focus on the occurrence of various superconducting phases 
is motivated by the standard BCS instability of the Fermi surface 
for arbitrarily weak particle-particle ($p$-$p$) interactions.
Under certain conditions, however, the particle-hole ($p$-$h$) channel 
might also become competitive. 
Here, a kinematic singularity in the corresponding Greens function 
only develops in (effectively) 1+1 dimensional systems, and at a 
total pair momentum of $Q=2p_F$ ($p_F$: Fermi momentum), known as Peierls 
instability~\cite{Pei55}. In higher dimensions it can nevertheless 
be relevant provided the interaction is strong enough.  One variant 
of $p$-$h$ instabilities are 'spin-density waves' as 
originally proposed by Overhauser~\cite{Ov62} for specific electronic 
materials (for a review see \cite{SDW_review}).
The analogue in the context of QCD, so called 'chiral density waves',  
has first been discussed by Deryagin~\etal~\cite{DGR92}. Using 
perturbative one-gluon exchange (OGE) at asymptotically high
densities it was shown that
the Overhauser-type pairing prevails over the BCS instability 
in the $N_c\to \infty$ limit ($N_c$: number of colors).
This is due to the fact that the BCS bound states, being  color non-singlet,
are dynamically suppressed by 1/$N_c$ as 
compared to the (colorless) Overhauser ones.
More recently, Shuster and Son~\cite{SS99} revisited  this mechanism 
for finite $N_c$ and including  Debye screening in the gluon 
propagator.  As a result, the chiral density wave  
dominates only for a very large number of colors, $N_c= {\cal O}(10^3)$.   
These findings have been confirmed in an analysis of coupled BCS/Overhauser
 equations using different arguments~\cite{PRWZ99}.
One concludes that  instabilities in the $p$-$h$ channel are not
relevant for real QCD at asymptotic densities.

The situation, however,  can be very different  
if the interaction strength between the quarks is substantially 
increased (to be referred to below as the strong coupling regime). 
A well-known example   
is the  Nambu-Jona Lasinio (NJL) description of chiral symmetry breaking  
in the  QCD vacuum (associated with the constituent quark mass 
gap and the build-up of the chiral condensate), which
requires a (minimal) critical coupling to occur.
At finite density, the same (attractive) interaction is operative
in the scalar-isoscalar $p$-$h$ channel. It's coupling strength is 
in fact augmented by a factor of $(N_c-1)$ over the (most attractive)
scalar diquark channel. On the other hand, geometric factors act
in its disfavor:  
unlike the BCS gap, which uniformly covers the entire Fermi surface,
the chiral density wave appears in form of `patches', 
their number depending on the symmetry of the presumed crystal.
The purpose of the present paper is to study the interplay
between Overhauser and BCS pairing, including different crystal structures,  
within the strong coupling regime. 
The focus is thus on quark matter at {\it intermediate}
densities,  \ie, large enough for the system to be in the quark phase,
but small enough to support nonperturbative interactions. 
This should roughly correspond to chemical potentials in the range 
$\mu_q\simeq$~0.4-0.6~GeV,  translating into baryon densities of 
3.5-12~$\rho_0$ 
(where $\rho_0$=0.16~fm$^{-3}$ denotes normal nuclear matter 
density).\footnote{The chiral crystal phase we are investigating is
not to be confused with another crystal phase discussed in 1980's 
related to $p$-wave pion condensation~\cite{Mig92} and later 
interacting skyrmions~\cite{sky}. 
Those works have addressed nuclear matter at lower densities, in which
the chiral condensate $<\bar q q>$ is only slightly perturbed from
its vacuum value and  basically uniform in space, while
the periodic structure is driven by pion fields.}

The article is organized as follows.
In Sect.~\ref{sec_gaps} we start by introducing the Nambu-Gorkov type
matrix propagator formalism that will subsequently be applied to obtain
the gap equations for the coupled BCS-Overhauser problem; special
attention is given to the single-quark spectra in the Overhauser ground 
state. In
Sect.~\ref{sec_results} we solve these equations using the
aforementioned variants of nonperturbative interactions, \ie,
somewhat schematic NJL-type as well as microscopic instanton-induced 
forces, for the slightly idealized case of two massless flavors and 
three colors.  In Sect.~\ref{sec_concl} we summarize
and discuss the relevance of our results for real QCD.

\section{Nambu-Gorkov Formalism and Coupled Gap Equations}
\label{sec_gaps}

\subsection{BCS Pairing}
A standard framework to address  multiple
instabilities in interacting many-body systems is provided by the 
Nambu-Gorkov formalism. Here, propagators are constructed as matrices
combining all potential condensate channels via off-diagonal 
elements (see, \eg, ref.~\cite{MJ68}), which automatically  
incorporates the interplay/coexistence of the various phases. 

For the familiar BCS case one adopts the following ansatz for the 
full propagator:
\beq
\hat{G}_{BCS}(k_0,\vec k,\Delta;\mu_q)
=\left( \ba{cc}
\la c_{k\uparrow} ~ c^\dagger_{k\uparrow} \ra  &
\la c_{k\uparrow} ~ c_{-k\downarrow}      \ra \\
\la c^\dagger_{-k\downarrow} ~ c^\dagger_{k\uparrow} \ra  &
\la c^\dagger_{-k\downarrow} ~ c_{-k\downarrow} \ra
               \ea \right) \
\equiv\left( \ba{cc}
G(k_0,\vec k,\Delta)  & \bar{F}(k_0,\vec k,\Delta)
\\
F(k_0,\vec k,\Delta) & \bar{G}(k_0,-\vec k,\Delta)
               \ea \right) \ .
\label{G_bcs}
\eeq
The gap equation is then derived by formulating the pertinent
Dyson equation,
\beq
\hat{G}_{BCS}=\left[ \hat{G}_0^{-1} -\hat{\Delta}\right]^{-1}=
\left( \ba{cc}
{G}_0^{-1}  &  \bar{\Delta} \\
  \Delta    & \bar{G}_0^{-1}
               \ea \right)^{-1} \ ,
\label{dy_bcs}
\eeq
which has the formal solution
\beq
\hat{G}_{BCS}=\frac{1}{G_0^{-1} \bar{G}_0^{-1} -\Delta\bar{\Delta}}
\left( \ba{cc}
\bar{G}_0^{-1}  &  -{\Delta} \\
  -\bar{\Delta} & G_0^{-1}
               \ea \right) \ ,
\label{sol_bcs}
\eeq
where
\beq
\Delta= (-i) \ \alpha_{pp} \int\frac{d^4p}{(2\pi)^4} \ F(p_0,\vec p,\Delta)
\label{gap_bcs}
\eeq
represents the (off-diagonal) 'selfenergy' contribution induced by
$p$-$p$ pairing (with an appropriate 4-fermion coupling constant
$\alpha_{pp}$), and
\bea
G_0 &=& \frac{1}{k_0-\epsilon_k+i\delta_{\epsilon_k}}
\\
\bar G_0 &=& \frac{1}{k_0+\epsilon_k+i\delta_{\epsilon_k}}
\eea
are the free particle propagator and its conjugate at finite
chemical potential (with $\epsilon_k=\omega_k-\mu_q$ and infinitesimal
$\delta_{\epsilon_k}=|\delta|~{\rm sgn}(\epsilon_k)$ according to
the sign of $\epsilon_k$).
Inserting the expression for the anomalous Greens function from 
eq.~(\ref{sol_bcs}),
\beq
F(k_0,\vec k,\Delta)=\frac{-\Delta}{(k_0-\epsilon_k+i\delta_{\epsilon_k})
(k_0+\epsilon_k+i\delta_{\epsilon_k})-\Delta^2} \ , 
\eeq
into the definition of $\Delta$, eq.~(\ref{gap_bcs}), yields
the gap equation. Notice that the pole structure of $F(k)$ always
ensures a nonvanishing contour for the energy integration.

\subsection{Overhauser Pairing}
On the same footing one can analyze pairing in the particle-hole
channel at finite total pair momentum $Q$. In the mean-field 
approximation (MFA)\footnote{This approximation is equivalent to the 
weak coupling approximation in band structure calculations where higher 
intra-band mixing is suppressed.}, the full Greens function and Dyson 
equation in the presence of a single stationary
wave take the form
\bea
\hat{G}_{Ovh}(k_0,\vec k,\vec Q,\sigma;\mu_q) &=& \left( \ba{cc}
\la c_{k\uparrow} ~ c^\dagger_{k\uparrow} \ra  &
\la c_{k\uparrow} ~ c^\dagger_{k+Q\downarrow}      \ra \\
\la c_{k+Q\downarrow} ~ c^\dagger_{k\uparrow} \ra  &
\la c_{k+Q\downarrow} ~ c^\dagger_{k+Q\downarrow} \ra
               \ea \right) \
\equiv\left( \ba{cc}
G(k_0,\vec k,\vec Q,\sigma)  & \bar{S}(k_0,\vec k,\vec Q,\sigma)
\\
S(k_0,\vec k,\vec Q,\sigma) & G(k_0,\vec k+\vec Q,\vec Q,\sigma)
               \ea \right)
\nonumber\\
 &=& \left[ \hat{G}_0^{-1}-\hat{\sigma}\right]^{-1} \ , 
\eea
which has the formal solution
\beq
S(k_0,\vec k,\vec Q,\sigma)=\frac{-\sigma}
{(k_0-\epsilon_k+i\delta_{\epsilon_k}) 
(k_0-\epsilon_{k+Q}+i\delta_{\epsilon_{k+Q}})-\sigma^2}
\label{prop_ovh}
\eeq
and gives the ensuing gap equation from the definition of the pairing
'selfenergy',
\beq
\sigma=(-i) \alpha_{ph}\int\frac{d^4p}{(2\pi)^4} \ 
S(p_0,\vec p,\vec Q,\sigma) \ .
\label{gap_ovh}
\eeq
Notice that here the energy contour integration receives nonvanishing
contributions only if
\beq
\epsilon_p \ \epsilon_{p+Q} - \sigma^2< 0 \ ,
\eeq
which means that the two poles in $p_0$ have to be in distinct 
(upper/lower) halfplanes,
\ie, one particle (above the Fermi surface) and one hole (below the
Fermi surface) are required to participate in the interaction. This
condition reflects on the particle-hole symmetry  caused by the nesting
of the Fermi surface in the presence of the induced 
wave. We stress again that  
an important difference to the BCS gap equation resides in the fact that
(for 2 or more spatial dimensions) one is not guaranteed a solution
for arbitrarily
small coupling constants since the $p$-$h$ Greens function $S$ does not
develop a kinematic singularity (as mentioned in the introduction 
this is very reminiscent to the QCD vacuum case of particle-antiparticle 
pairing across the Dirac sea).

At finite densities, the
formation of a condensate carrying nonzero total momentum $Q$
is associated with nontrivial spatial structures, \ie, crystals,
characterized by a 'lattice spacing' $a=2\pi/Q$. In three dimensions 
a more complete
description thus calls for the inclusion of additional wave vectors.
In general, the $p$-$h$ pairing gap can be written as
\beq
\sigma(\vec{r})=\sum\limits_j \sum\limits_{n=-\infty}^{+\infty}
\sigma_{j,n} e^{in\vec{Q}_j\cdot \vec{r}} \ ,
\eeq
where the $\vec{Q}_j$ correspond to the (finite) number of
fundamental waves, and the summation over $|n|>1$ accounts for
higher harmonics in the Fourier series.
The matrix propagator formalism allows for the treatment of
multiple waves through an  expansion of the
basis states according to
\beq
\hat{G}=\left( \ba{cccc}
\la c_{k\uparrow} ~ c^\dagger_{k\uparrow} \ra        &
\la c_{k\uparrow} ~ c^\dagger_{k+Q_x\downarrow} \ra  &
\la c_{k\uparrow} ~ c^\dagger_{k+Q_y\downarrow} \ra  &
 \cdots                                              \\
\la c_{k+Q_x\downarrow} ~ c^\dagger_{k\uparrow} \ra        &
\la c_{k+Q_x\downarrow} ~ c^\dagger_{k+Q_x\downarrow} \ra  &
\la c_{k+Q_x\downarrow} ~ c^\dagger_{k+Q_y\downarrow} \ra  &
 \cdots                                              \\
\la c_{k+Q_y\downarrow} ~ c^\dagger_{k\uparrow} \ra        &
\la c_{k+Q_y\downarrow} ~ c^\dagger_{k+Q_x\downarrow} \ra  &
\la c_{k+Q_y\downarrow} ~ c^\dagger_{k+Q_y\downarrow} \ra  &
 \cdots                                              \\
\vdots & \vdots & \vdots &\ddots  \\

\ea
\right)  \ .
\label{G_ovh}
\eeq
In practice the expansion has to be kept finite.
The possibility of additional BCS pairing is straightforwardly incorporated
into eq.~(\ref{G_ovh}) by extending the latter with the off-diagonal
states from eq.~(\ref{G_bcs}).

In what follows we will consider up to $n_w=6$ waves
in three orthogonal directions with $Q_x=Q_y=Q_z$ and $n=\pm 1$,
characterizing a cubic crystal through three standing waves with
the fundamental modes (for simplicity we will also assume the magnitude
of the various Overhauser condensates to be equal, \ie,
$\sigma_j\equiv \sigma$). The important new features that arise
through introducing additional states become already apparent in the
simplest extension to 2 condensates. In this case one has for the
(coupled) gap equation(s)
\bea
\sigma_x &=& (-i) \ \alpha_{ph} \int\frac{d^4p}{(2\pi)^4} \
 \frac{-\sigma_x \ G_0^{-1}(\vec p+Q_y)} 
{G_0^{-1}(\vec p) \ G_0^{-1}(\vec p+Q_x) \ G_0^{-1}(\vec p+Q_y)-
\sigma_x^2 \ G_0^{-1}(\vec p+Q_y) - \sigma_y^2 \ G_0^{-1}(\vec p+Q_x)} 
\nonumber\\
 &=& (-i) \ \alpha_{ph} \int\frac{d^4p}{(2\pi)^4} \  
 \frac{-\sigma_x \ G_0(\vec p+Q_x)} {G_0^{-1}(\vec p)-
\sigma_x^2 \ G_0(\vec p+Q_x) - \sigma_y^2 \ G_0(\vec p+Q_y)}
\label{sig_x}
\eea
(and an equivalent one for $\sigma_y$ by interchanging $x\leftrightarrow y$).
For finite $\sigma_y$ additional possibilities for the nonvanishing of
the energy contour integration appear through an extra zero
of the third-order polynomial in the denominator of the full propagator 
in eq.~(\ref{sig_x}).
This enlarges the integration region and can be interpreted as
interference effects between the patches (or waves).
Diagrammatically
this can be understood as an additional insertion of the $\sigma_y$
condensate on a  particle (or hole) line of energy $\epsilon_p$.
Note that a priori it is not clear whether 
such interferences are constructive or
destructive, that is, give a positive or negative contribution to
the right-hand-side (\rhs) of eq.~(\ref{sig_x}).

In the propagators $G_0$ contributions from antiparticles have been
neglected. This should be a reasonable approximation in the
quark matter phase at sufficiently large $\mu_q$, \ie, after the 
usual (non-oscillating particle-antiparticle) chiral
condensate has vanished. At the same time, since we base our analysis
on nonperturbative forces, the
applicable densities are bounded from above. Taken together, we estimate 
the range of validity for our calculations to be roughly given by
0.4~GeV~$\lsim \mu_q \lsim$~0.6~GeV. This coincides with
the regime where, for the physical current strange quark mass of
$m_s\simeq 0.14$~GeV, the two-flavor superconductor might prevail
over the color-flavor locked (CFL) state so that our restriction
to $N_f=2$ is supported.

\subsection{Spectrum in the Overhauser Case}
The poles of the mean-field propagators discussed above provide the
quasiparticle excitations in both the BCS and Overhauser case. In the
former, the spectrum consists of gapped particles and holes. In the
latter case, the physical interpretation is rendered more subtle by the 
presence of a standing wave. To keep the analysis transparent, we will 
discuss analytic results in 1+1 dimension and proceed
to a numerical evaluation in 3+1 dimensions.

In 1+1 dimension, the quasi-particle excitations following from the pole
condition for the propagator in the Overhauser case, eq.~(\ref{prop_ovh}),
have energies
\beq
\epsilon_\pm = \frac 12 (\epsilon_k + \epsilon_{k-Q})
\pm \sqrt{(\epsilon_k -\epsilon_{k-Q} )^2 +\sigma^2}
\label{SPEC1}
\eeq
with $\epsilon_k=|k|-\mu_q$ and $Q=2\pi/a$. This spectrum can be 
understood if we note that the quarks are moving in a self-induced 
potential $V(x)=-2\sigma\, {\rm cos} \, (Qx)$ through the stationary
wave. Indeed, in the presence of such a potential the spectrum is
banded with $|k|\leq \pi/a$ representing the first Brillouin Zone ($BZ$-1). 
In weak coupling the spectrum is mostly free except at $k=0,\pm \pi/a$
where band-mixing is large. For $\mu_q < Q$ we can ignore most 
of the band mixing except for the lowest one near the edge of $BZ$-1. 
Degenerate perturbation theory gives readily
\bea
\left| \ba{cc}
k_0 -\epsilon_k& \sigma
\\ \sigma & k_0 -\epsilon_{k-Q}
               \ea \right| = 0\,\,,
\label{DET1}
\eea
in agreement with (\ref{SPEC1}). The $2\times 2$ character of (\ref{DET1})
follows from mixing between two bands. As foot-noted above, 
it is analogous to the MFA where the band-mixing is 
treated in the extended description commonly used in weak-coupling 
band-structure calculations.

\bef[t]
\bce
\epsfig{file=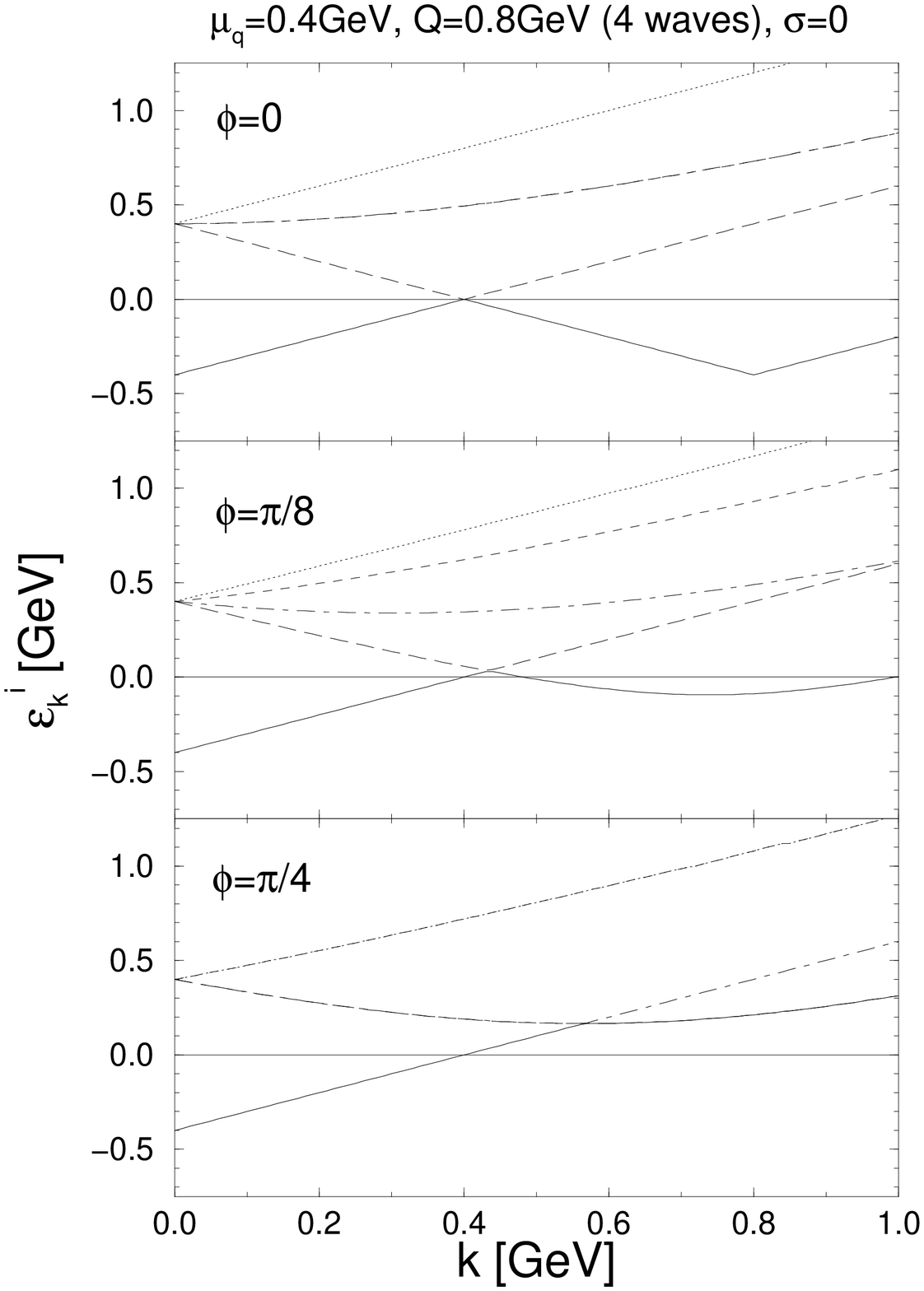,width=7.8cm}
\hspace{0.5cm}
\epsfig{file=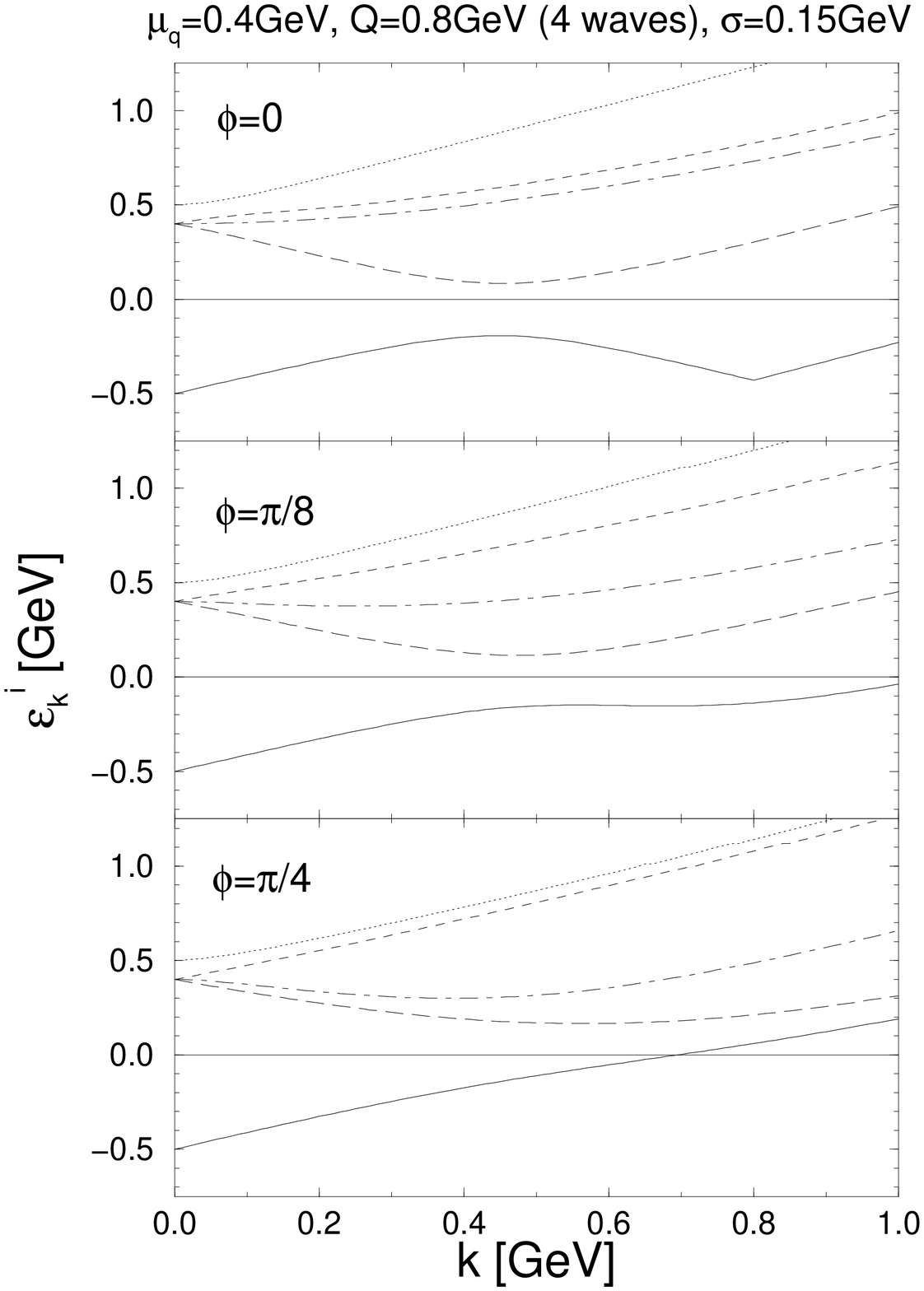,width=7.8cm}
\ece
\caption{Dispersion relations of the various particle and hole
branches using 4 waves in $\pm x$ and $\pm y$ direction with equal
gaps and wave vector moduli $Q=2p_F$. The solutions are displayed
for 3 angles of the momentum with respect to the $x$-axis in the
$x$-$y$ plane. By symmetry, the solutions repeat itself within
each octant (\ie, every $45^o$). The left panel shows the noninteracting
case ($\sigma=0$), and the right panel has been obtained by setting
$\sigma=0.15$~GeV.}
\label{fig_disp800}
\eef

The quasiparticles of energy $\epsilon_-$ are characterized by a standing wave
$\psi_- (x) \approx {\rm cos} \,(\pi x/a)$, and those of energy $\epsilon_+$
are characterized by a standing wave $\psi_+ (x)\approx {\rm sin}(\pi x/a)$
near the edge of the Brillouin zone. The energy is
substantially lowered by the standing wave $\psi_-(x)$ with a probability
density in opposite phase to the potential. The standing wave $\psi_+ (x)$
corresponds to a probability density in phase with the potential, hence
substantially more expensive energetically. At the edge of the zone, the
two states are gapped by $2\sigma$. Clearly, the lowest energy state is
reached by filling only those states corresponding to $E_-$, that is by
setting the Fermi energy at the gap. The ensuing state is an insulator.

In higher dimensions, the band-mixing becomes more intricate. However, 
in the weak-coupling approximation and for Fermi momenta in the vicinity 
of $Q/2$, higher intra-band mixing is small and we may just use the 
extended band-structure description which is equivalent to our 
mean-field treatment. The quasiparticle spectra $\epsilon_k^j$ 
($j=0, \dots, n_w$) 
follow numerically from the poles of the propagator.\footnote{Again, 
$n_w$ refers to the number of plane waves retained in $\sigma (\vec{x})$ 
with $n_w/2$ being typically the number of standing waves.}
Fig.~\ref{fig_disp800} shows an example of 4 waves in $\pm x$ and
$\pm y$ directions for the canonical value of the wave vector, $Q=2p_F$,
with ($\sigma_x=\sigma_y>0$) and without ($\sigma_x=\sigma_y=0$) interactions.
One clearly recognizes the formation of the gap close to the degeneracy
point (level crossing) in the non-interacting case, which happens in the
vicinity of the Fermi surface. For $Q=p_F$
(fig.~\ref{fig_disp400}) energy
gain arises from an appreciable pushing down of the lowest level which lies
rather deep within the Fermi sea. This is (partially) counteracted by an 
upward push of the upper branch which also corresponds to occupied states.   
Thus one expects the most beneficial  
configuration to be when the Fermi surface lies in between split levels.  
\bef
\bce
\epsfig{file= 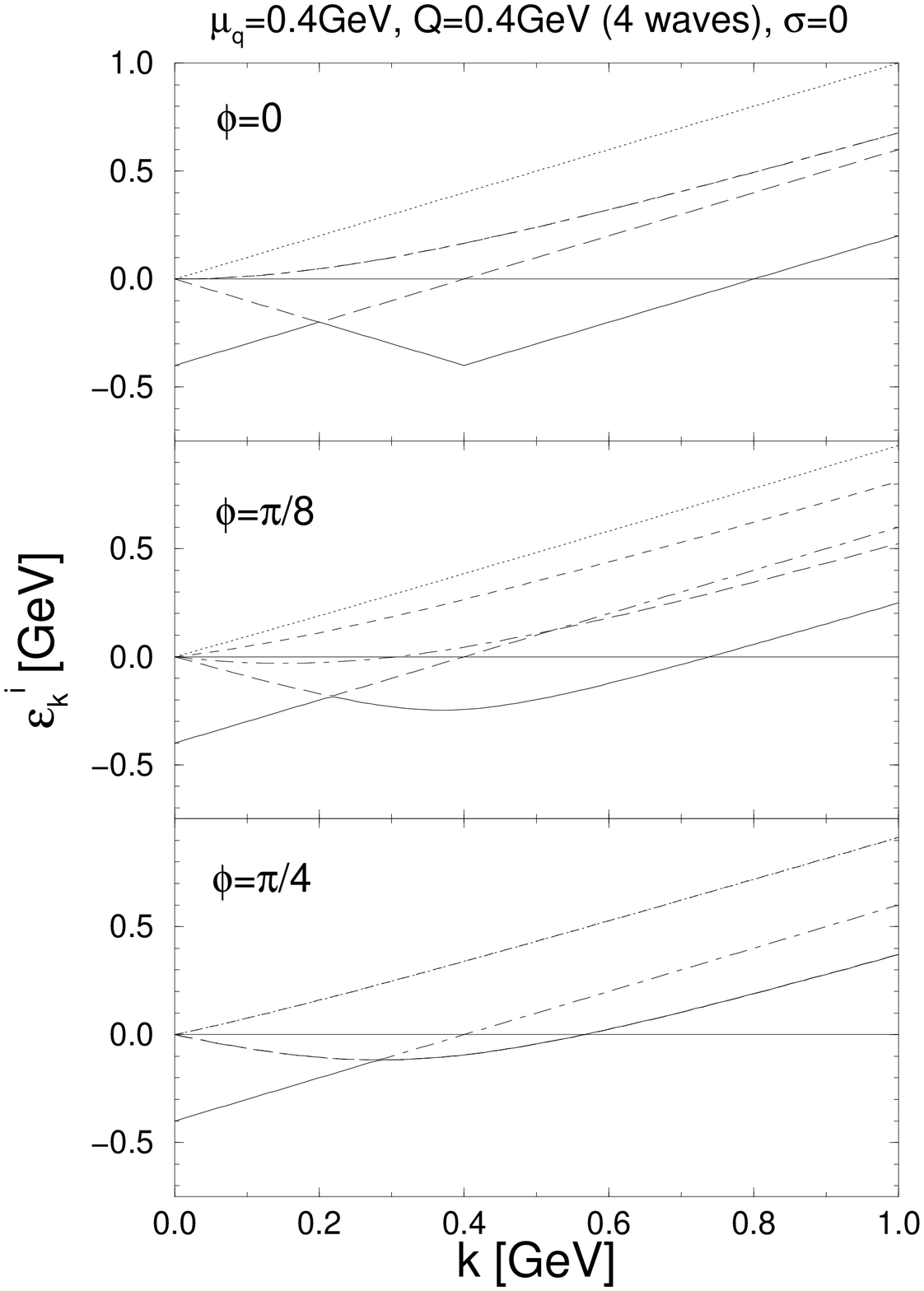,width=7.8cm}
\hspace{0.5cm}
\epsfig{file= 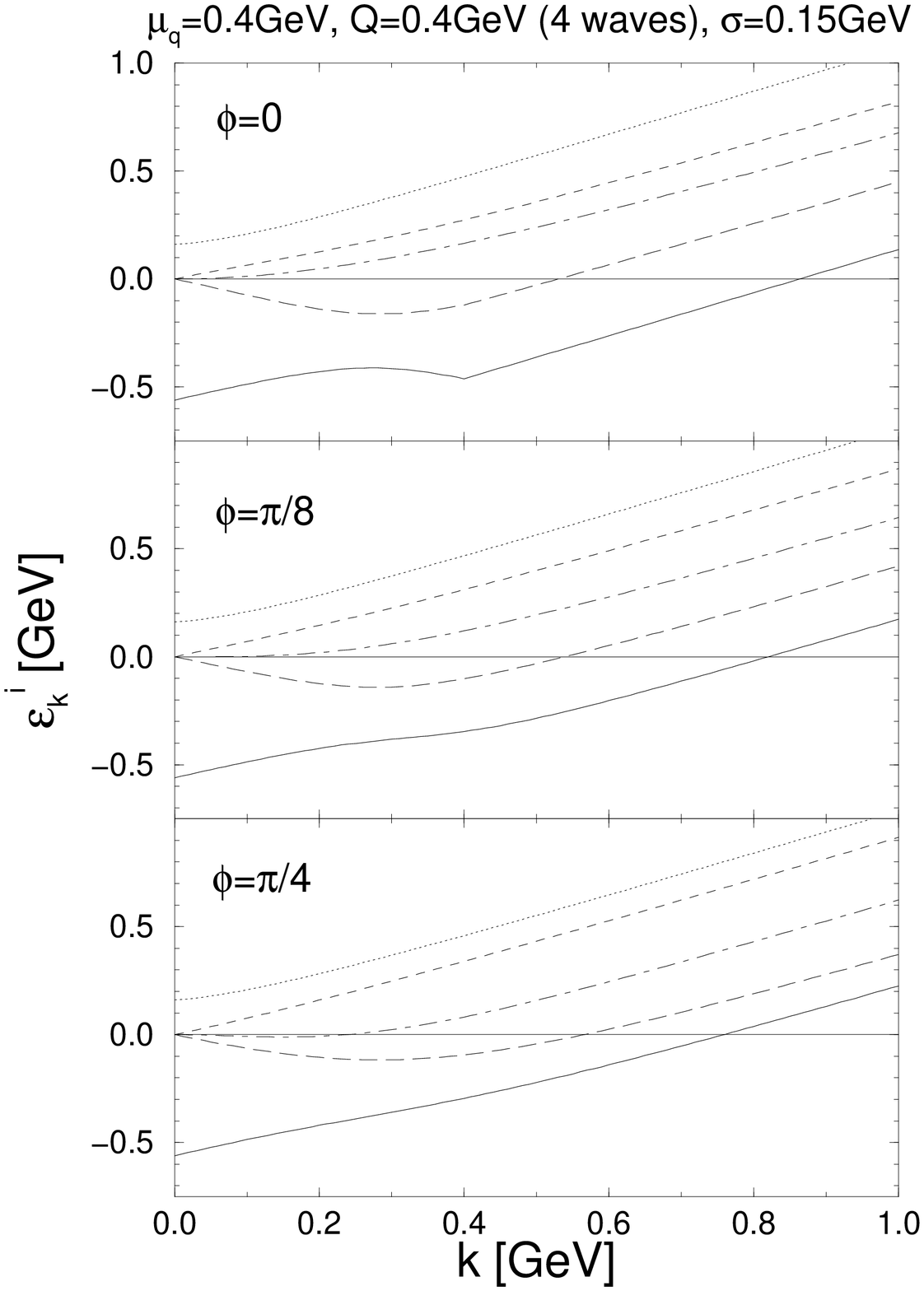,width=7.8cm}
\ece
\caption{Same as fig.~\protect\ref{fig_disp800} but for
$Q=p_F$.}
\label{fig_disp400}
\eef

\subsection{Energy Budget and Periodicity}
Solutions of the gap equations correspond to extrema (minima) in the
energy density with respect to (\wrt) the gap $\sigma$. However, solutions
may exist for several values of the wave vector $Q$. To determine the
minimum in the latter quantity, one has to take recourse to the explicit 
form of the free energy density. In MFA,
\bea
V_3\,\Omega (\mu_q, Q, \sigma) = 
\int\, d^3x \,\left(
 \frac{\sigma^2 (x)}{2\lambda} +
\left<q^\dagger \, ( i{\alpha}\cdot\nabla -2\sigma (x)\, q) \right>\right)\,\,,
\eea
where $V_3$ is the 3-volume. 
The first contribution removes the double counting from the fermionic 
contribution in the mean-field treatment.  Retaining only the particle 
contribution (\ie, neglecting antiparticles), the free energy simplifies 
to 
\bea
\Omega(\mu_q,Q,\sigma) &=&
\Omega_{pot}(\mu_q,Q,\sigma) +\Omega_{kin}(\mu_q,Q,\sigma)
\nonumber\\  
&=&  \sum\limits_{j=1}^{n_{w}} \left[
\frac{C(N_c,N_f)}{\lambda} \  \sigma_j^2 \right] +
\sum\limits_{j=0}^{n_{w}}
2 N_c N_f \int\limits_{BZ-1} \frac{d^3k}{(2\pi)^3} \ \epsilon_k^j 
\ \Theta(-\epsilon_k^j) \ 
\label{omtot}
\eea
with color/flavor coefficients ${C(N_c,N_f)}$ which will
depend on the concrete form of the pairing interaction. To avoid 
double counting the integration for the kinetic energy part is restricted 
to $BZ$-1 as defined by the momentum regions
$[-\vec Q_j/2,\vec Q_j/2]$ (as well as $|\vec k|\le k_F$). 
This amounts to a folding of the various 
branches into $BZ$-1 and enforces the explicit lattice
periodicity onto the free energy. 
This point is illustrated in fig.~\ref{fig_bzones} for the free case 
($\sigma\equiv 0$) along one spatial direction. For fixed chemical
potential smaller wave vectors require the inclusion of an increasing number
of branches to correctly saturate the available states within the Fermi
sea (through a multiple folding until the Fermi surface is reached).
{\it E.g.}, for $\mu_q \le 3Q/2$ the lowest two harmonics with 
$k_x\pm Q_x$ suffice. Above, the next two higher harmonics with 
$k_x\pm 2Q_x$ are necessary to encompass the occupied energy states within 
$BZ$-1 up to $\mu_q\le 5Q/2$, and so forth. When using additional
waves in other spatial directions similar criteria hold (albeit more 
complicated due to nontrivial angular dependencies).     
\bef
\vspace{-0.9cm}
\bce
\epsfig{file=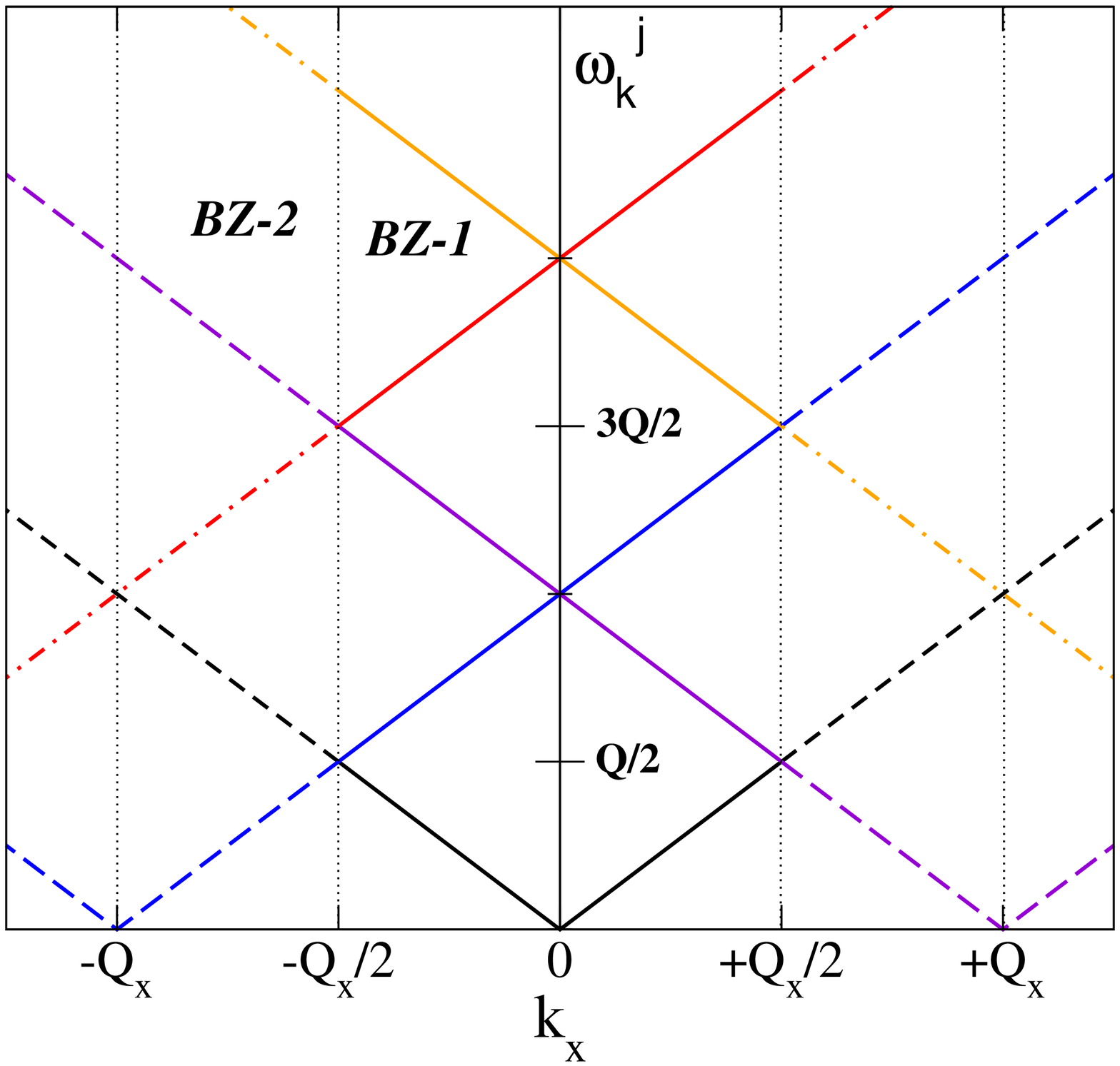,width=8.5cm,angle=-90}
\ece
\vspace{-0.2cm}
\caption{Free quasiparticle dispersion relations ($\sigma_{j}\equiv 0$) for 
massless quarks in a crystal with periodicity in x-direction. The 
vertical dotted lines indicate the boundaries of the first ($BZ$-1) and 
second ($BZ$-2) Brillouin Zone (corresponding to $|k_x|\le Q_x/2$ 
and $Q_x/2\le |k_x|\le Q_x$, respectively). Plotted are the 5 branches 
$\omega_k^j =|k_x|$, $|k_x\pm Q_x|$ and $|k_x\pm 2Q_x|$ with the full lines 
characterizing their contributions to $BZ$-1.}
\label{fig_bzones}
\eef

Solutions of the gap equations in general support different pairs of 
$\{\sigma,Q\}$; the combination that minimizes $\Omega(\mu_q,Q,\sigma)$ 
is the thermodynamically favored one.
Note that the gap equations are not subject to explicit momentum restrictions
since off-shell momenta of arbitrary magnitude can in principle contribute.

As a simple example, (\ref{omtot}) can be explicitly computed in
1+1 dimensions by recalling that the Fermi surface coincides with the gap,
\ie, $\omega_{k_F}=\mu_q=k_F=Q/2$. Specifically, the contribution from the 
Fermi sea is
\bea
\Omega_{kin} = \rho\, \epsilon_F \, 
\left( 1-\frac 12 \, \left( \sqrt{1+\xi^2} +\xi^2\,{\rm ln}\, 
(\xi + \sqrt{1+\xi^2} ) \right) \right) \ ,
\eea
with $\xi=\sigma/Q$ and a density $\rho=dk_F/\pi$ where $d$ is the overall
degeneracy. In MFA, the induced standing wave is 
$\sigma (x) = 2\sigma\, {\rm cos} (Qx)$, and the double counting in 
the Fermi sea is removed by
\bea
\Omega_{pot} =   \frac 1L \int\, dx \, \frac {\sigma^2 (x)}{2\lambda}
=\frac {\sigma^2}{\lambda}\,\,.
\eea
The minimum value of $\sigma$ can be obtained in this case analytically
by minimizing $\Omega =\Omega_{kin}+\Omega_{pot}$.

\section{Nonperturbative Forces and Results}
\label{sec_results}

For the actual calculations we need to specify the quantum
numbers of the pairing channels. To do so we
take guidance from low- (or zero-) density
phenomenology encoded in effective 4-point interactions.
In the particle-hole channel the strongest attraction is in the
'$\sigma$' channel given by (including exchange terms~\cite{RSSV00})
\beq
{\cal L}_{mes}^{\sigma}=\frac{\lambda}{8 N_c^2} \ (q\bar q)^2 \ ,
\label{L_sig}
\eeq
whereas in the particle-particle channel it is believed to
be the scalar diquark in the color-antitriplet channel (which, in fact,
arises from a Fierz-transformation of (\ref{L_sig})),
\beq
{\cal L}_{diq}^{\bar 3}=\frac{\lambda}{8 N_c^2 (N_c-1)}
(q^TC\gamma_5\tau_2\lambda_A^a q) \
(\bar q\tau_2 \lambda_A^a\gamma_5C\bar{q}^T) \
\label{L_diq}
\eeq
($C$: charge conjugation matrix, $\lambda_A^a$: antisymmetric
color matrices, $\tau_2$: SU(2)-flavor matrix). For practical use
the effective vertices have to be supplemented with
ultraviolet cutoffs. In the following we will consider two variants 
thereof and discuss the pertinent results for the coupled
Overhauser/BCS equations.

\subsection{NJL Treatment}
\label{sec_njl}
In a widely used class of Nambu-Jona Lasinio models the ultraviolet
behavior of the pointlike vertices is regulated by
3- or 4-momentum multipole formfactors (or even sharp $\Theta$-functions).
We here employ a dipole form,
\beq
F(p)=\left(\frac{\nu \Lambda^2}{\nu \Lambda^2+p^2}\right)^\nu
\label{ff_njl}
\eeq
($\nu=2$), for each in- and outgoing quark line with $\Lambda=0.6$~GeV
as a typical 'chiral' scale (variations within such parametrizations 
do not affect our qualitative conclusions in this section).
The coupling constant $\lambda=67$~fm$^2$ is calibrated to a constituent
quark mass of $M_q=0.4$~GeV in vacuum. Note that there is no well-defined
way of introducing density dependencies into the interaction.
Since at finite $\mu_q$ the relevant quark interactions occur 
at the Fermi surface, a formfactor of type (\ref{ff_njl}) implies
the loss of interaction strength with increasing $p_F$.

This schematic treatment has been shown to yield robust results
for 2-flavor BCS pairing with gaps $\Delta\simeq 0.1$~GeV at
quark chemical potentials around 0.5~GeV~\cite{ARW98}. Including
now the $p$-$h$ pairing as outlined in the previous section we find only  
rather fragile evidence for the emergence of chiral density waves (at
$\mu_q=0.4$~GeV): 
for wave vectors $Q_x\le 0.150$~GeV the \rhs~of the Overhauser gap 
equation  supports solutions with gaps around $\sim 5$~MeV. The smallness 
of $Q_x$ in fact requires six waves (with $k\pm nQ_x$, $n$=1,2,3) to fill
all states within the Fermi sphere. Somewhat more robust solutions are 
obtained when increasing the 4-fermion coupling constant. 
{\it E.g.}, with
a vacuum constituent quark mass of $M_q=0.5$~GeV, which implies 
$\lambda=73$~fm$^2$, the minimum solution emerges for $Q_x\simeq 0.2$~GeV 
and $\sigma\simeq 20$~MeV.  
However, the gain in the total free energy is very small: 
$\Omega=-1.2973*10^{-3}$~GeV$^{-4}$
as compared to the free Fermi gas value of 
$\Omega=-1.2969*10^{-3}$~GeV$^{-4}$
(to be contrasted with the BCS ground state for which 
$\Omega_{BCS}(\mu_q=0.4~{\rm GeV})\simeq -1.375*10^{-3}$~GeV$^{-4}$ 
at a pairing gap of $\Delta\simeq 0.13$~GeV). 

We also checked that the incorporation 
of waves in other spatial directions does not lead to 
further energy gain.


\subsection{Instanton Approach at Finite Chemical Potential}
\label{sec_inst}
A more microscopic origin of effective 4-fermion interactions
is provided within the instanton framework. In the
finite-density context it has previously been employed to study
the competition between the chiral condensate and two-flavor
superconducting quark matter in refs.~\cite{CD99,RSSV00}.
Let us briefly recall some elements of the approach.
The starting point is the QCD partition function in instanton
approximation,
\beq
{\cal Z}_{inst}(\mu_q)=\frac{1}{N_+!~N_-!}
\prod\limits_{I=1}^{N_+,N_-}
\int d\Omega_I \ n(\rho_I) \ e^{-S_{int}^{gluon}}
\left[ det(i\not{\!\!D}-i\mu_q\gamma_4)\right]^{N_f} \ ,
\eeq
where $\Omega_I=\{\Theta_I,\rho_I,z_I\}$ denote the collective coordinates 
(color, size and position) of the instanton solutions and $n(\rho_I)$ their 
individual weight.
To extract effective quark interactions one reintroduces quark fields
in a way that is compatible with the fermionic determinant of the previous
equation~\cite{DP86},
\beq
{\cal Z}_{inst}(\mu_q)=\int {\cal D}\psi {\cal D}\psi^\dagger
\exp\left[\int d^4x \psi^\dagger (i\not\!\partial-i\mu_q\gamma_4)\psi\right]
\int \frac{d\lambda_\pm}{2\pi}
\exp\left[\lambda_\pm Y_\pm
+N_\pm\left(\ln\left[\frac{N_\pm}{\lambda_\pm V_4}\right]-1\right)\right] \ ,
\eeq
and an additional auxiliary integration over $\lambda_\pm$ has been
introduced to exponentiate the effective (2$N_f$)-fermion vertices $Y_\pm$.
For two flavors the latter are given by
\bea
\lambda_\pm \ Y_\pm &=& \lambda_\pm
\int \frac{d^4k_1 d^4k_2 d^4p_1 d^4p_2}{(2\pi)^{16}} \
(2\pi)^4 \delta^{(4)}(k_1+p_1-k_2-p_2)  \
\nonumber\\
& & \qquad\qquad\qquad\qquad \times
\left[\psi^\dagger{\cal F}^\dagger(p_1,-\mu_q) \gamma_\pm \tau_a^-
{\cal F}(k_1,\mu_q) \psi\right] \
\left[\psi^\dagger{\cal F}^\dagger(p_2,-\mu_q) \gamma_\pm \tau_a^-
{\cal F}(k_2,\mu_q) \psi\right] \
\eea
with $\gamma_\pm=(1\pm\gamma_5)$, flavor matrices $\tau_\alpha=(\vec\tau,i)$
and the instanton formfactors
${\cal F}(p,\mu_q)=(p+i\mu_q)^-\varphi(p,\mu_q)^+$,
which are matrices in Dirac space, adopting the notation of ref.~\cite{CD99},
\ie, $x^\pm\equiv x_\mu\sigma_\mu^\pm$ with
$\sigma_\mu^\pm\equiv(\pm\vec\sigma,1)$. Since the fermionic determinant
has been approximated by its zero-mode part, the formfactors are entirely
determined by the Fourier-transformed quark zero-mode wave functions,
\bea
\phi_{I,A}(p,\mu_q)
&=& \int d^4x \ e^{-ip\cdot x} \ \phi_{I,A}(x,\mu_q)
\nonumber\\
&=& \varphi(p,\mu_q)^\pm \ \chi_{R,L}
\eea
with $\phi(x,\mu_q)$ satisfying the Dirac equation in the background
of an (anti-) instanton,
\beq
\not\!\!D_{I,A} \ \phi_{I,A}(x)=0 \ .
\eeq
The explicit form of $\varphi_\mu(p,\mu_q)$ can be found in
refs.~\cite{CD99,RSSV00}.

As before we preselect  the potential condensation channels
as  the scalar-isoscalar $p$-$h$ and $p$-$p$ ones
which, after solving the (matrix) Dyson equation, yields the coupled
gap equations
\bea
\Delta &=&  \frac{(-i) \ \lambda}{N_c (N_c-1)}
\int  \frac{d^4p}{(2\pi)^4} \ B(p;\mu_q) \ F(p;\Delta,\sigma_j,Q_j)
\label{gapbcs}
\\
\left(\sigma_x+5~\delta\sigma_x\right) &=& 
\quad \frac{(-i) \ \lambda}{N_c} \quad
\int  \frac{d^4p}{(2\pi)^4} \ A(p,Q_x;\mu_q) \
\ S_x(p;\Delta,\sigma_j,Q_j)
\label{gapovh}
\\
\left(\sigma_{3,x} -10~\delta\sigma_x\right) &=& 
\quad \frac{(-i) \ \lambda}{N_c} 
\quad \int  \frac{d^4p}{(2\pi)^4} \ A(p,Q_x;\mu_q) \
\ S_x(p;\Delta\equiv 0,\sigma_{3,j},Q_j) \ .
\label{gapovh3}
\eea
Here, $\delta\sigma_j=\sigma_j-\sigma_{3,j}$ denotes the difference in 
the Overhauser gaps for quarks of color 1,2 ($\sigma_j$) or 
color 3 ($\sigma_{3,j}$) 
which, respectively, do (eq.~(\ref{gapovh})) or do not (eq.~(\ref{gapovh3})) 
participate in the diquark pairing (once $\Delta\ne 0$)~\cite{CD99,RSSV00}.    
In principle, the wave vectors in the color-1,2 and color-3 channels 
could also be different when minimizing the free energy. However, 
the actual solutions for $\sigma_j$ and $\sigma_{3,j}$ turn out  
to be very close to each other even
in the presence of large BCS gaps $\Delta$ so that the generically very
smooth dependence on the $Q_j$ should not cause appreciable deviations 
between the two color sectors.  
Under our simplifying assumption that the momentum moduli $|Q_j|$ (as well
as the associated gap parameters $\sigma_j$) of the Overhauser pairing
are of equal magnitude the additional $(2n_w-2)$ gap equations 
for the other $p$-$h$
channels are equivalent to (\ref{gapovh}) and (\ref{gapovh3}).
The explicit form of the propagators is given by
\bea
S_x(p,\Delta,\sigma_j,Q_j) &=& -A(p,Q_x;\mu_q) \ \sigma_x(p,Q_x) \
 G_0(p+Q_x) \ D(p,\Delta,\sigma_j,Q_j)
\\
F(p,\Delta,\sigma_j,Q_j) &=& -B(p;\mu_q) \ \Delta(p) \
 \bar{G}_0(p) \ D(p,\Delta,\sigma_j,Q_j)
\eea
with
\beq
D(p,\Delta,\sigma_j,Q_j)=\bigl[\Delta(p)^2 \bar{G}_0(p) - G_0^{-1}(p)
           +\sum\limits_j \sigma_j(p,Q_j)^2 \ G_0(p+Q_j) \bigr]^{-1} \ . 
\eeq
The functions $A$ and $B$ represent the (square of the) instanton form
factors (normalized to one in vacuum) acting on each fermion line
entering/exiting a vertex, and we have introduced the notation
$\sigma_j(p;Q_j)\equiv \sigma_j A(p;Q_j)$, $\Delta(p)=\Delta B(p)$.
The integration variable $\lambda\equiv\lambda_\pm$ plays
the role of an effective coupling constant, which, however, is not a priori
fixed. Rather, its value is found from minimization of the
free energy via a saddle point condition, which reads~\cite{CD99}
\bea
\frac{N}{V}&=&\lambda \ \langle Y_+ + Y_-\rangle  \ ,
\nonumber\\
&=& \frac{1}{\lambda} \
[2N_c^2 \sum\limits_j {\sigma_j}^2 +4N_c(N_c-1)\Delta^2] \ ,
\label{lambda}
\eea
where $\langle Y_+ + Y_-\rangle$ denotes the ground state expectation
value of the interaction vertices with potential condensates.
Thus the magnitude of the gaps itself governs the effective coupling
to the instantons. In the present treatment the instanton density $N/V$ 
is assumed to be constant.\footnote{This assumption is motivated by the 
observation that the free energy associated with the instanton vacuum
(background) is rather large compared to interaction corrections
arising in the finite-density quark sector. It is corroborated by 
explicit calculations in the 'cocktail model' of 
ref.~\cite{RSSV00}, where the grand potential has been minimized 
explicitly over both the properties of the instanton ensemble as well 
as quark Fermi sphere: the resulting variations in the total $N/V$ 
where found to be at the few percent level. However, the assertion of 
constant $N/V$ can imply unphysical behavior when the system is driven  
towards very large or very small pairing gaps.}     
The final result for the free energy at the minimum
then becomes
\beq
\Omega(\mu_q)=-\frac{\ln{\cal Z}}{V}=\Omega_{kin}(\mu_q)+\frac{N}{V}
\ln\left[\frac{\lambda}{(N/V)}\right] \ ,
\label{ominst}
\eeq
indicating that the potential (second) term favors small values for 
$\lambda$, whereas the kinetic (first) term exhibits the usual decrease
with increasing values for the gaps/condensates (and thus for 
$\lambda$). We should also point out that $\Omega(\mu_q)$ is only
determined up to an overall constant which is associated with the
nonperturbative vacuum energy of about $-0.5$~GeV/fm$^3$ (or, equivalently, 
bag pressure $P$$>$0). This term is encoded in the scale dependence
of the argument in the logarithm (note that $\lambda$ and $N/V$ have
different dimensions), which we have not assessed here since it is 
not relevant for our analysis (this will be the origin of positive 
values for $\Omega(\mu_q)$ encountered below). 

Before we come to the numerical solutions of the gap equations let us
recall the specific density dependence of the instanton
formfactors, cf. fig.~\ref{fig_ffinst}. At fixed energy (upper panel)
the strength of the interaction is clearly concentrated at the Fermi
surface: the falloff with three-mometum sets in
only above $p\simeq p_F$. On the other hand, as a function
of (Euclidean) energy the strength is reduced starting from $p_4=0$ (see
lower panel).
These features reflect that the instanton zero-modes, which
mediate the interaction, operate  across
the Fermi surface, \ie, at zero energy but at 3-momenta equal
to the Fermi momentum. This behavior is quite distinct from the schematic
(density-independent) NJL forces employed in the previous section.
Already at this point one can anticipate the Overhauser pairing 
to be more competitive than in the NJL treatment.
\bef
\vspace{-1.0cm}
\bce
\epsfig{file=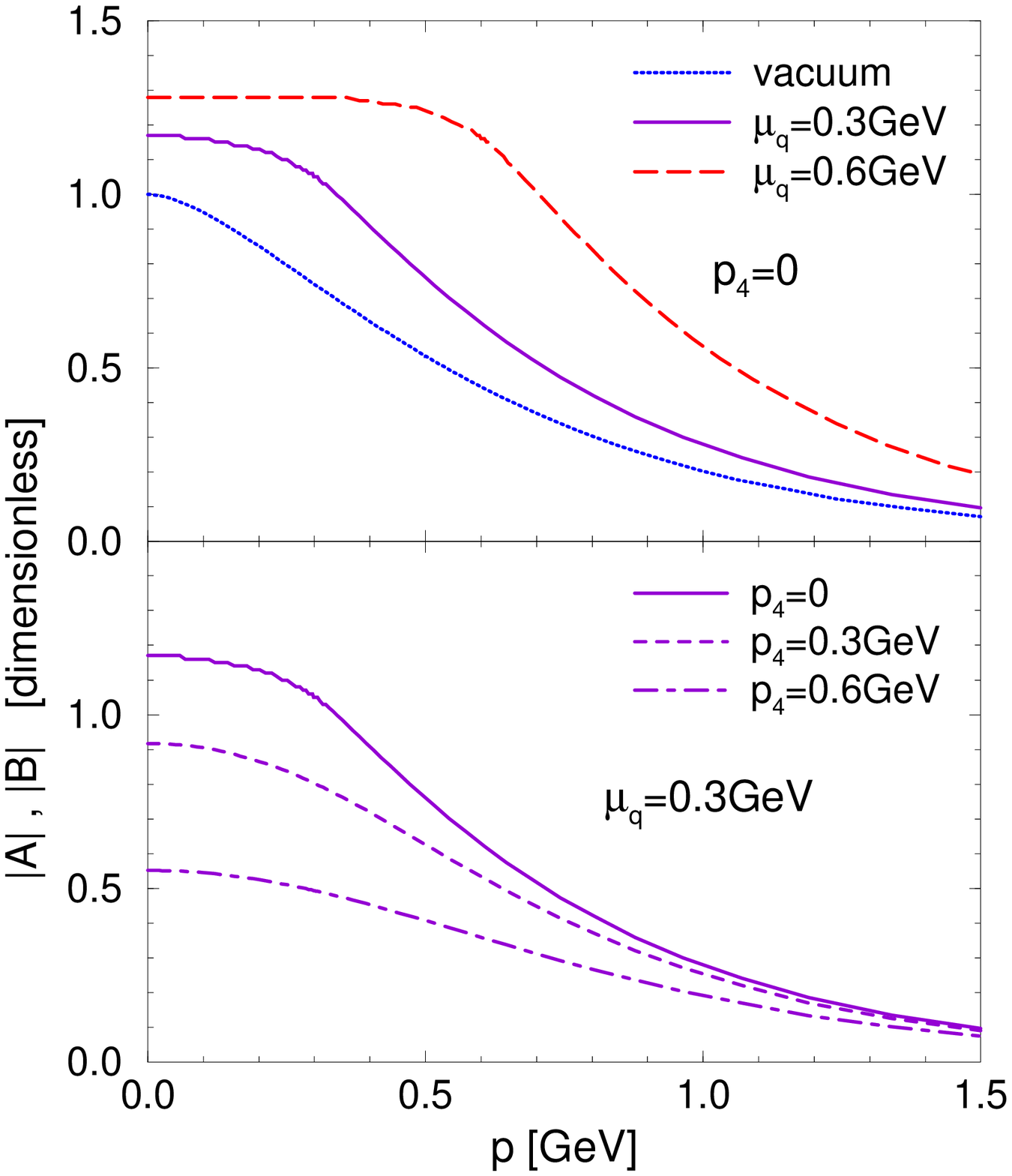,width=9.6cm}
\ece
\vspace{-2.5cm}
\caption{In-medium instanton form factors as a function
of three-momentum. Upper panel: at fixed energy $p_4=0$
for chemical potetnials $\mu_q=0$, 0.3~GeV, 0.6~GeV;
lower panel: for fixed $\mu_q=0.3$~GeV and various energies.}
\label{fig_ffinst}
\eef

For the evaluation of the kinetic part of the free energy as given
in eq.~(\ref{omtot}) a complication arises from the fact that the 
the instanton formfactors are defined in Euclidean space.  
We therefore approximated the gaps entering into the 
integral for $\Omega_{kin}$ by their zero-energy values retaining the 
3-momentum dependence, \ie, $\Delta(p)\simeq \Delta(\vec p) =
\Delta B(p_4=0,\vec p)$,  and equivalently for $\sigma$. It turns out that
this approximation is consistent in the sense that the resulting 
extrema in $\Omega$ are in line with the solutions of the gap equations 
(which are solved in Euclidean space with the full 4-momentum
dependence of the formfactors). Other choices for  fixing $p_4$ do not
comply with this criterium.

If not otherwise stated, the subsequent calculations have been performed 
for a total instanton density of $N/V=1$~fm$^{-4}$ which in vacuum translates 
into  a constituent quark mass of $M_q=0.34$~GeV.
The search for simultaneous solutions to the gap equations 
(\ref{gapbcs}), (\ref{gapovh})  and (\ref{gapovh3}) together with the 
selfconsistency condition on the coupling, eq.~(\ref{lambda}), is 
illustrated in fig.~\ref{fig_rhs}. 
The upper two curves represent the values for $\sigma$ and $\sigma_3$ 
that simulatneuosly solve the two Overhauser gap eqs.~(\ref{gapovh}) and 
(\ref{gapovh3}) at given BCS gap $\Delta$ (plotted on the abscissa).
The lower curve indicates the values for $\sigma$ that solve the BCS
gap eq.~(\ref{gapbcs}) for given $\Delta$ (using the value for $\sigma_3$ 
from the upper curve to fix the coupling $\lambda$). Thus a coexistence
state of Overhauser and BCS pairing would be signalled by the crossing
of the two lower lines.  As mentioned above no such state is found;
the only physical solutions correspond to the crossing points
of the upper (full or dashed) curve with the $y$-axis 
(Overhauser state with $\Delta=0$,
$\sigma=\sigma_3$ finite) and of the lower (dashed-dotted) curve with 
the $x$-axis 
(BCS state with $\sigma=\sigma_3=0$ and $\Delta=0.225$~GeV; the discrepancy of 
about 15\% (less at higher $\mu_q$) with refs.~\cite{CD99,RSSV00} reflects 
the accuracy when neglecting antiparticle states).
\vspace{-0.7cm}
\bef
\bce
\epsfig{file=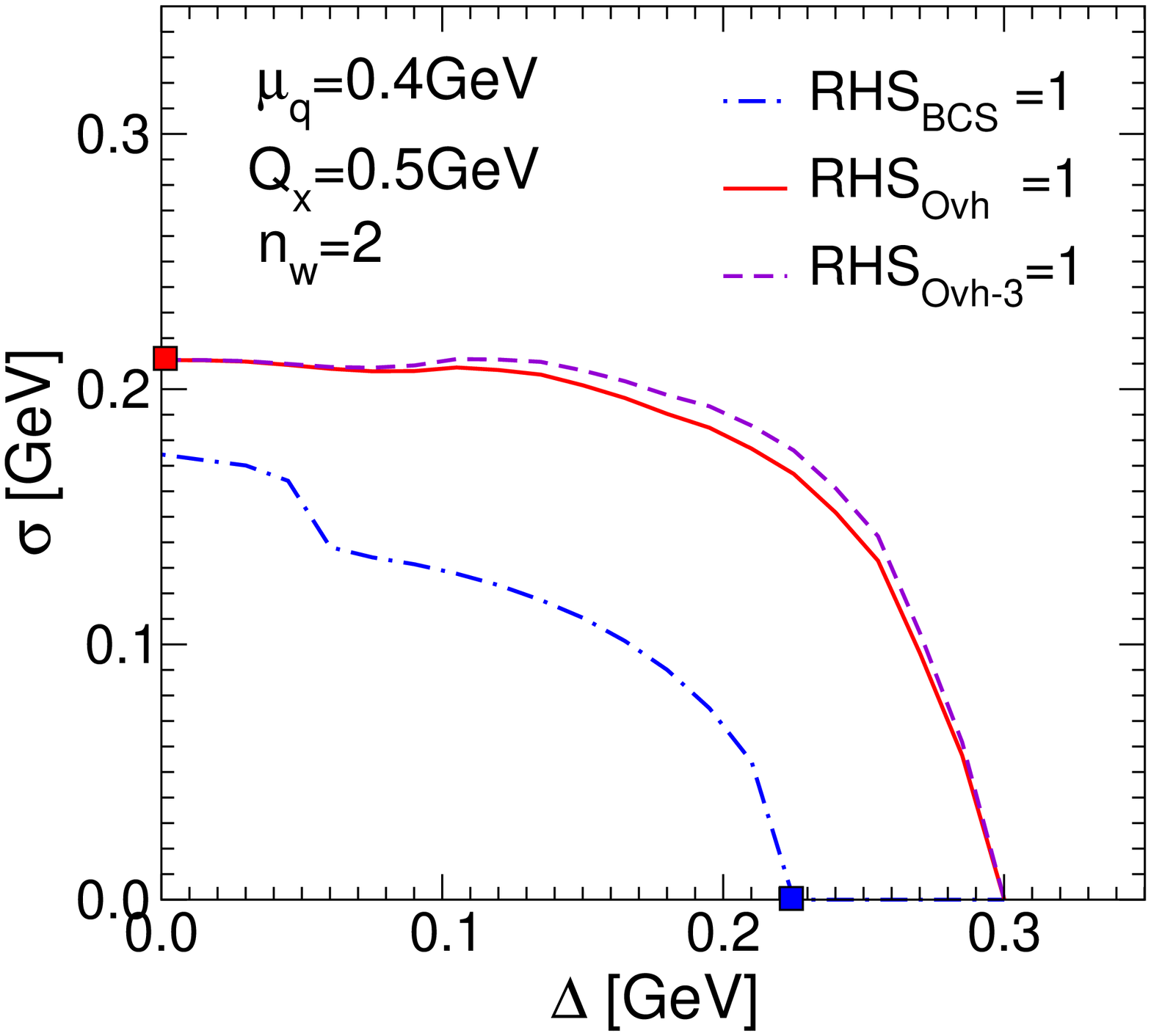,width=8cm,angle=-90}
\ece
\caption{Solutions of the coupled Overhauser and BCS equations
with 2 waves at  $Q_x=\pm 0.5$~GeV.
The full and dashed lines indicate, respectively, the values for 
$\sigma_x(\Delta)$ and $\sigma_{3,x}(\Delta)$ which simultaneously
solve the two Overhauser gap equations (\protect\ref{gapovh}) and 
(\protect\ref{gapovh3}) for given $\Delta$. 
The dashed-dotted line corresponds to the points $\sigma_x,\Delta$
that solve the BCS gap equation (\protect\ref{gapbcs}).
A nontrivial simultaneous solution to all 3 gap equations would be
signalled by a crossing of the full and dashed-dotted curves. The two
independent pure BCS and Overhauser solutions are marked by the squares
on the $x$- and $y$-axis, respectively.}
\label{fig_rhs}
\eef

The question then is which one is thermodynamically favored. The BCS
solution is unique ($\Delta=0.225$~GeV) and gives a total free
energy of $\Omega_{BCS}(\mu_q=0.4~{\rm GeV})=2.3*10^{-3}$~GeV$^4$ 
(up to a constant which is not relevant here, as discussed above).

The situation is more involved for the Overhauser configurations.  
Let us start with the 'canonical' case where the momentum vector of the
chiral density waves is fixed at twice the Fermi momentum, $Q=2p_F$.   
In fig.~\ref{fig_Ndep} the resulting minimized free energy 
(corresponding to solutions of eq.~(\ref{gapovh})) is displayed as function
of the number of included waves. The Overhauser solutions are not far  
above the BCS groundstate, with a slight energy gain for 
an increased number of waves. 

However, one can further economize the energy of the Overhauser state
by exploiting the freedom associated with the wave vector $Q$ (or,
equivalently, the periodicity of the lattice).
For $Q>2p_F$ the free energy rapidly increases. On the other hand,
for $Q<2p_F$ more favorable configurations are found. To correctly
assess them one has to
include the waves in pairs $|k\pm Q_j|$ of standing waves
(\ie, $n_w=2,4,6,\dots$)
to ensure that the occupied states in the Fermi sea are saturated
within the first Brillouin Zone (cf.~fig.~\ref{fig_bzones}).
The lowest-lying state we could find at $\mu_q=0.4$~GeV occurs for one
standing wave with $Q_{min}\simeq 0.5$~GeV and $\sigma\simeq 0.21$~GeV
with a free energy $\Omega\simeq 2.3*10^{-3}$~GeV$^4$, practically
degenerate with the BCS solution. In solid state physics the breaking of
translational invariance in one spatial direction is typically associated
with 'liquid crystals', \ie, 2-dimensional layers of (uniform) liquid
separated by periodic spacings of $a=2\pi/Q$.
In our case, $a\simeq 2.5$~fm. The minimum in the wave vector is
in fact rather shallow, as seen from the explicit momentum dependence
of the free energy displayed in fig.~\ref{fig_Qdep}.
\bef
\vspace{-0.7cm}
\bce
\epsfig{file=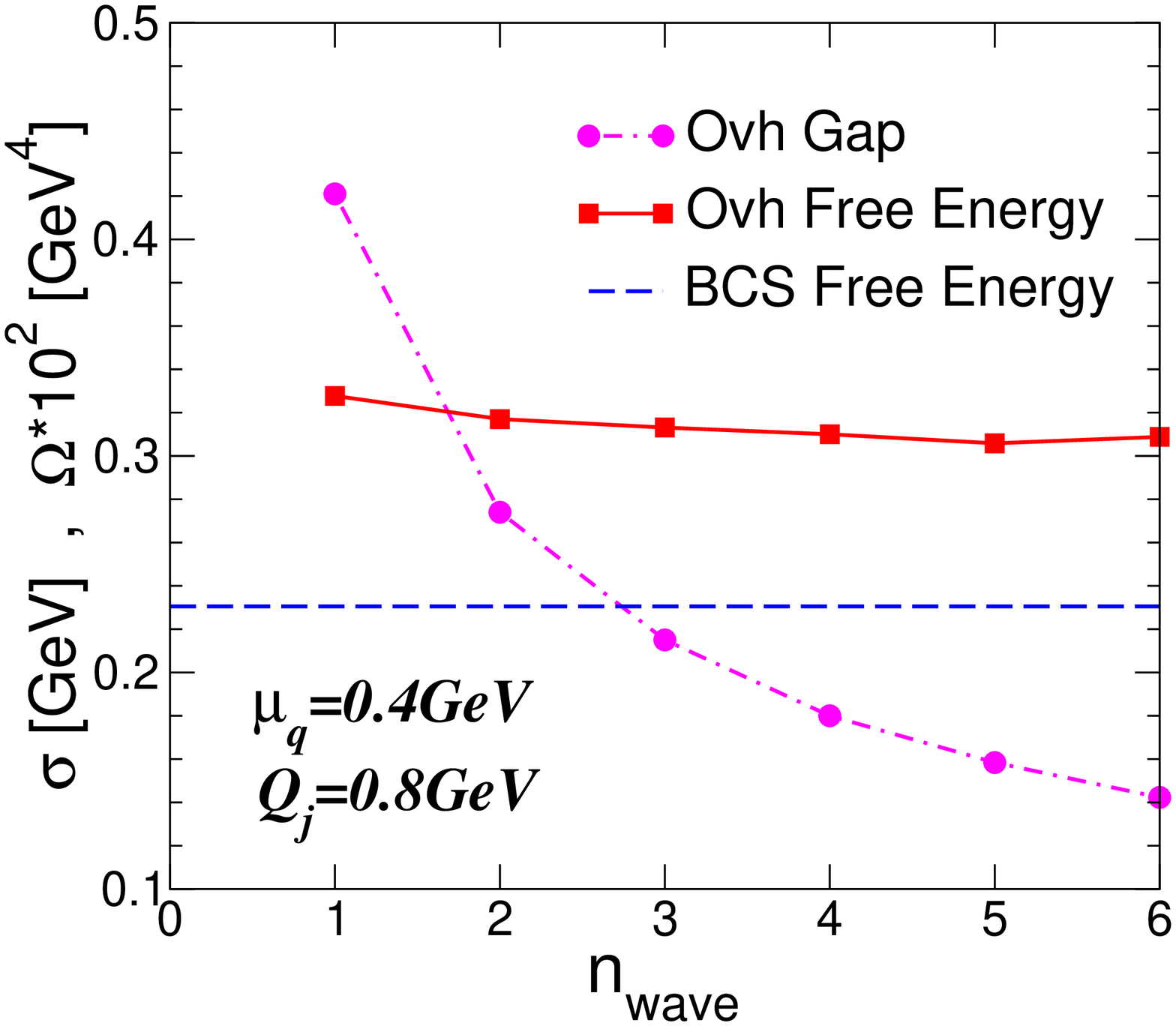,width=8cm,angle=-90}
\ece
\vspace{-0.3cm}
\caption{Dependence of the free energy (upper full line) and associated 
$p$-$h$ pairing gap (dashed-dotted) on the number of included
waves ('patches') with fixed 3-momentum modulus $|\vec Q_j|=0.8$~GeV
for solutions of
eqs.~(\protect\ref{gapovh}) and (\protect\ref{lambda}); the full line
marks the value of the BCS ground state free energy that solves
eqs.~(\protect\ref{gapbcs}) and (\protect\ref{lambda}). The results
are for $\mu_q=0.4$~GeV and $N/V=1$~fm$^{-4}$.}
\label{fig_Ndep}
\eef
\bef
\vspace{-0.7cm}
\bce
\epsfig{file=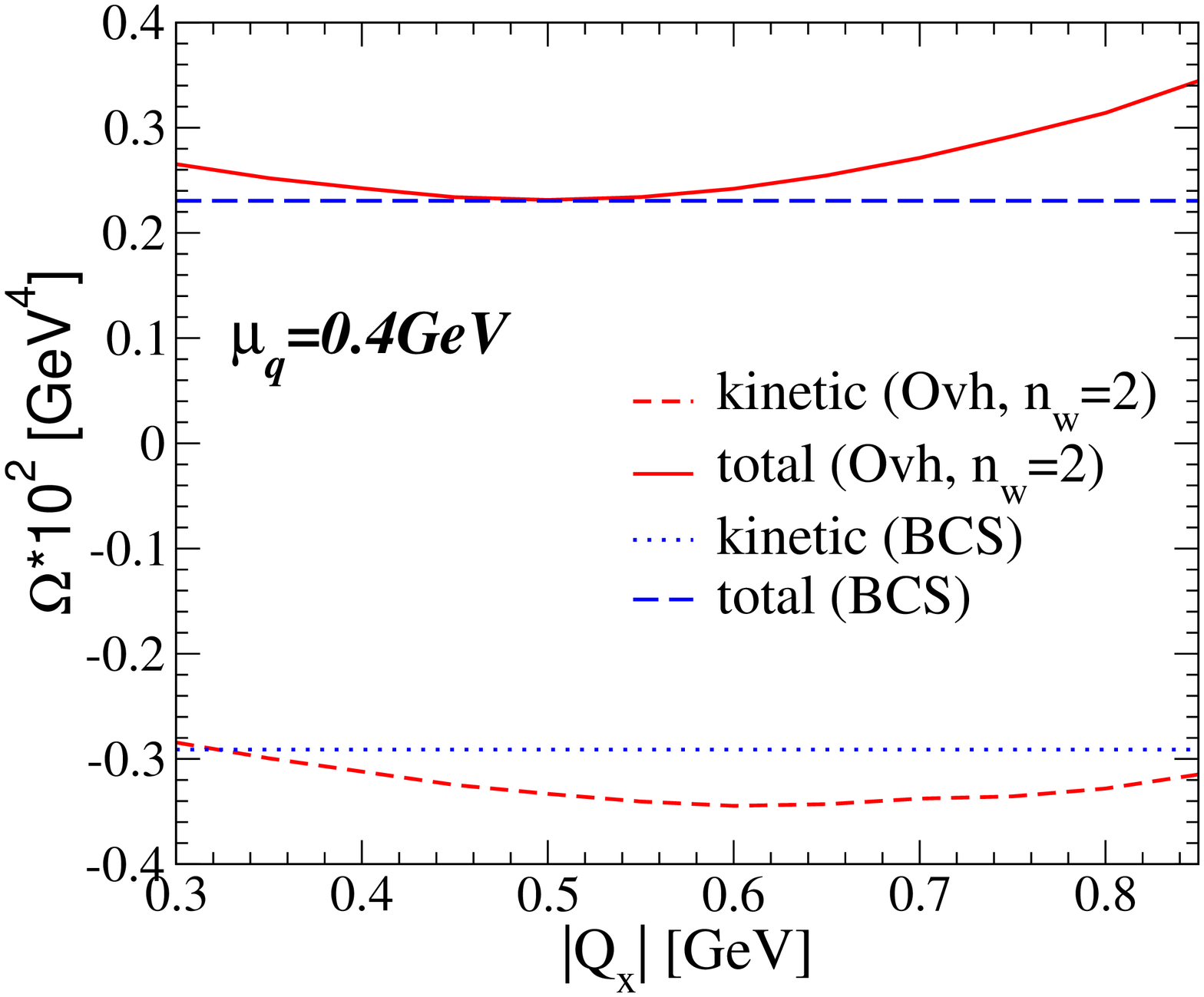,width=7.3cm,angle=-90}
\hspace{-1.4cm}
\epsfig{file=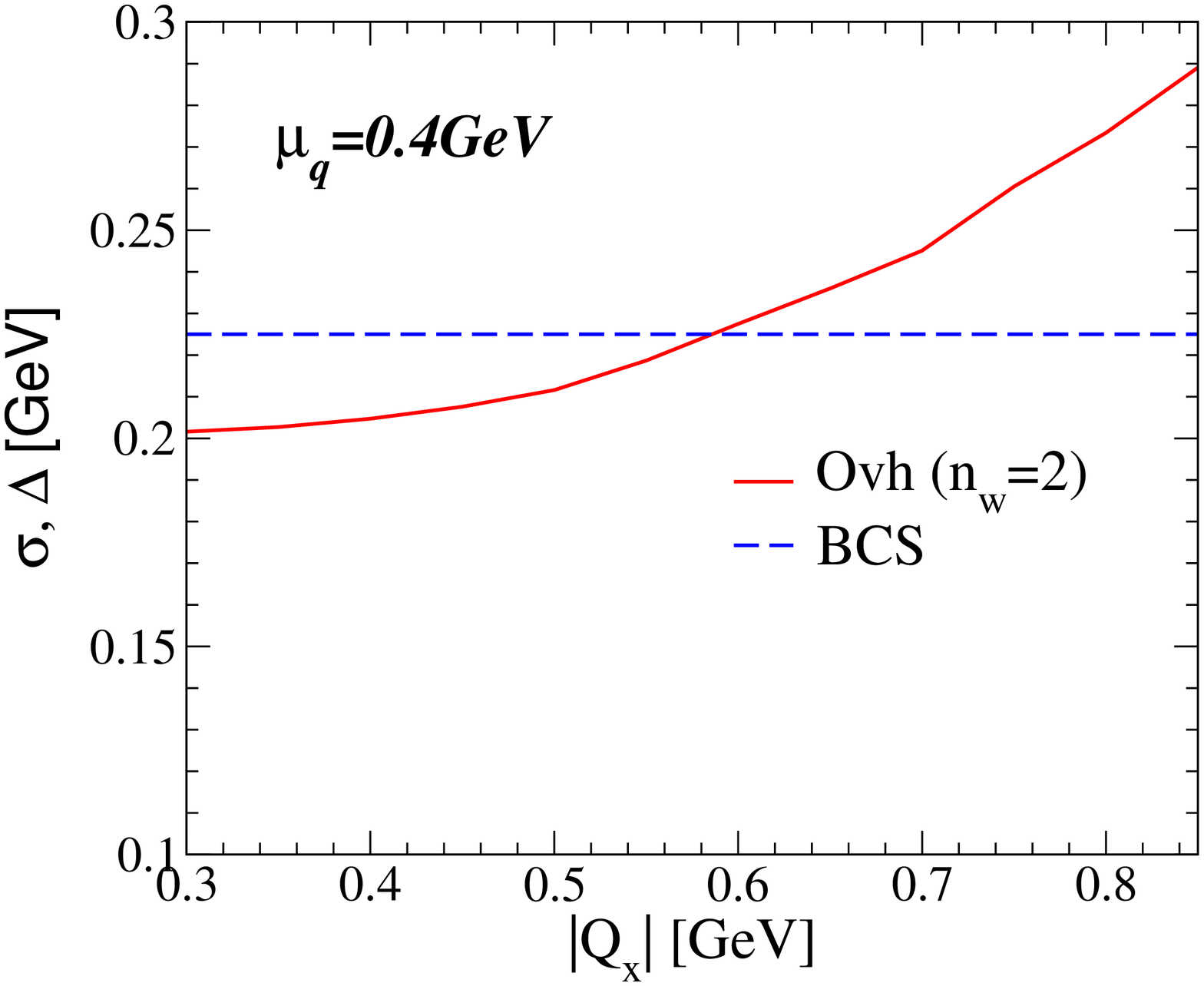,width=7.3cm,angle=-90}
\ece
\vspace{-0.3cm}
\caption{Left panel: wave-vector dependence of the Overhauser free energy 
for one standing wave  (full line: $\Omega_{tot}^{Ovh}$, short-dashed line: 
$\Omega_{kin}^{Ovh}$) in comparison to the BCS solution 
(long-dashed line: $\Omega_{tot}^{BCS}$, dotted line: $\Omega_{kin}^{BCS}$)
at $\mu_q=0.4$GeV (we note again that 
the absolute values of the total free energies are only determined up to an 
overall (negative) constant (related to the 'bag presssure') 
which drops out in the relative comparison of the solutions);
right panel: wave-vector dependence of the Overhauser pairing gap (full line)
compared to the BCS gap (long-dashed line).}
\label{fig_Qdep}
\eef

Finally we confront  in fig.~\ref{fig_MUdep} the density dependence
of the free energies (upper panel) and pairing gaps (lower panel) in
the minimum of the BCS and $n_w$=2,6 Overhauser states. 
Again we see that over the applicable $\mu_q$-range the solutions are 
close in energy, with (almost) degenerate minima for the 
$n_w$=2-'liquid crystal'  and the BCS ground state at the lower 
densities of  $\rho_B\simeq 4\rho_0$.  
Towards higher densities, where the gaps and thus the strength of the 
effective instanton interactions decrease, the BCS solution becomes 
relatively  more favorable.
This confirms once more that in the 3+1 dimensions the Overhauser pairing
can only compete for sufficiently strong coupling.  
We should also note that for both the $n_w$=2- and $n_w$=6-case as displayed 
the optimal momentum vector of the pertinent standing waves stays at 
approximate values of $Q_j\simeq 5p_F/4$ and   $5p_F/3$, respectively.  
\bef
\bce
\epsfig{file=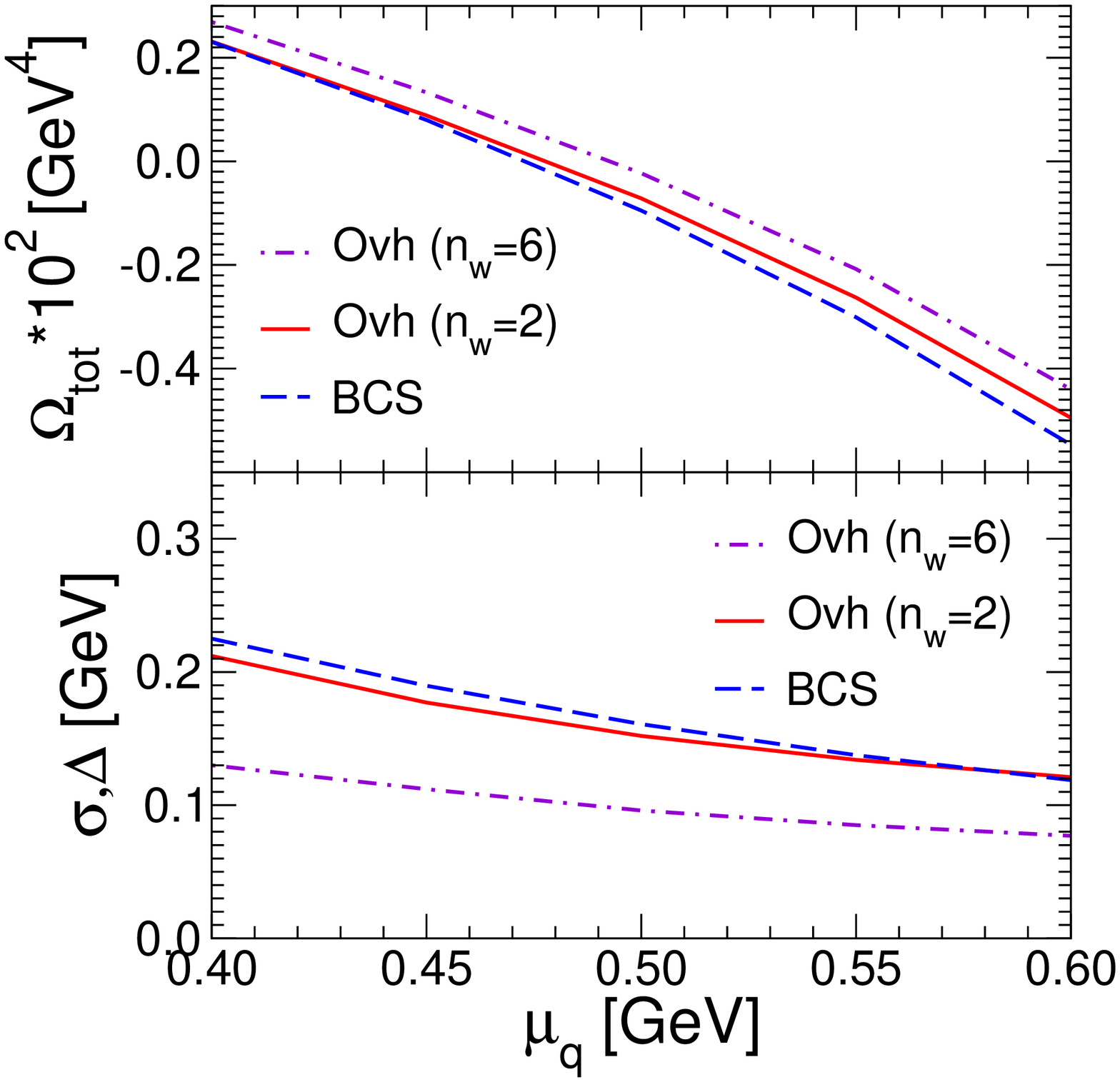,width=10cm,angle=-90}
\ece
\caption{Chemical potential dependence of the 
total free energies (upper panel) and pairing gaps (lower panel) 
for the BCS (long-dashed lines) and Overhauser (solid and dashed-dotted 
lines) solutions.}
\label{fig_MUdep}
\eef

\section{Concluding Remarks}
\label{sec_concl}
Employing a standard  Nambu-Gorkov (matrix) propagator approach
we have performed an analysis of competing instabtilities
in the particle-particle and particle-hole channel for three-color,
two-flavor QCD at moderate quark densities. As an essential ingredient
we used nonperturbative forces (strong coupling) and preselected the
potential condensation channels with guidance from low-energy
hadron phenomenology, \ie,  both the diquark as well as the quark-hole
pairing were evaluated in their scalar-isoscalar channels.
The corresponding coupled gap equations do not seem do support
simultaneous nontrivial solutions. Needless to say that our
calculations might be further improved by including higher standing waves
through a larger class of crystalline symmetries (polyhedron
structures). Also, our mean-field approximation may be extended
to account for higher intra-band mixings when $2\mu_q>Q$.

This not withstanding, an important outcome of our
calculation is that the individual (separate) solutions for the BCS
and Overhauser ground states are quite close in energy, indicating the
importance of particle-hole instabilities when using interaction
strengths as typical for low-energy hadronic binding. This
becomes particularly evident when comparing schematic NJL interactions
with microscopic instanton forces: whereas the former
are reduced in magnitude with increasing density (entailing a relative 
suppression of the Overhauser pairing), the density-dependent instanton
form factors essentially preserve their strength at the Fermi surface
rendering the Overhauser pairing competitive. Indeed,
with a somewhat larger instanton-density of $N/V=1.4$~fm$^{-4}$,
corresponding to a vacuum constituent quark mass of $M_q=0.41$~GeV,
the Overhauser solution reaches below the BCS one.
On the other hand, if we were to minimally account for
strange quarks, the instanton interaction (being a 6-quark vertex)
in the $ud$ sector would lose about 60\% of its strength due
to the reduced strange-quark mass ($m_s\simeq$~0.15~GeV as opposed to
$M_s\simeq0.45$~GeV in vacuum) in closing-off the strange quark line.
Some of this loss might be recovered once strange quarks themselves  
participate in the Overhauser pairing.
Our findings are to be contrasted with earlier calculations based on
perturbative OGE at high densities where an extremely large number of
colors was required for the Overhauser pairing (\ie, a chiral density wave) 
to overcome the BCS instability.

Finally, a remark about the relevance of our results for neutron stars
is in order. Here, quark matter is believed to reside mostly in a mixed phase,
with significant charge separation between quark and nuclear components.
The quark core, however, also exhibits charge asymmetry due to the  
finite strange quark mass. Consequently, quark matter  should have 
appreciably different up- and down-quark chemical  potentials,
which suppresses the flavor-singlet diquark ($ud$) pairing~\cite{Beda}. 
On the other hand, such
a suppression does not apply to the (flavor-singlet) particle-hole 
channels of type $uu^{-1}$, $dd^{-1}$, which suggests 
that isopin asymmetric quark matter
provides additional favor to the chiral crystal.

\section*{ACKNOWLEDGMENTS}
We thank M. Alford, G. Carter, K. Rajagopal and A. Wirzba for
useful discussions. 
This work was supported by DOE grant No. DE-FG02-88ER40388.


\end{document}